\documentclass[prb,amsmath,amssymb,showpacs,twocolumn]{revtex4}
\usepackage{graphicx}
\def\be{\begin{equation}}
\def\ee{\end{equation}}
\def\Ic{I_{{\rm c}0}}
\begin{document}
\title{Noise rectification by a superconducting loop with two weak links}
\author{Jorge Berger}
\affiliation{Physics Unit, Ort Braude College, P. O. Box 78,
21982 Karmiel, Israel and \\
Department of Physics, Technion, 32000 Haifa, Israel}
\email{phr76jb@tx.technion.ac.il}
\begin{abstract}
We consider a superconducting loop with two weak links that encloses a magnetic
flux. The weak links are unequal and are treated
as Josephson junctions with non-sinusoidal phase dependence. We devise a model
that takes into account the fluctuation of the critical currents, due to the
fluctuations of the order parameter in the weak links. These fluctuations are
important near the onset of superconductivity; in this regime they may significantly
weaken and eventually disconnect the superconducting loop.
As a consequence of these fluctuations and of the resistive
noise in the junctions, the average dc voltage does not vanish. Our model can be
easily extended to provide a qualitative description of a recent experiment.
\end{abstract}
\maketitle

\section{INTRODUCTION}
This study is a cross of two well established phenomena: Brownian motors
and fluctuations near a phase transition. The offspring of this combination will
be a circuit that can rectify thermal noise and sustain a dc voltage.

Brownian motors are systems in which thermal noise leads to the diffusion of some 
coordinate (typically, the position of a particle), whereas a time-dependent applied
potential (typically, periodic in some direction of space) confines this coordinate and controls how far and when it is allowed to diffuse. If the applied potential is asymmetric,
then the fluctuating coordinate drifts in a preferred direction and Brownian motion is rectified. Many reviews\cite{astum1,Rei,astum2,RMP,Zim} and experiments\cite{exp} on
Brownian motors are available.

Brownian motors are also called ``flashing ratchets", alluding to the fact that the
confining potential pulsates in time. A related phenomenon is a ``rocked ratchet";
in this case the confining potential is static, but an additional time-dependent applied potential is superposed to it. This addition distorts the confining potential in such a 
way that any of its minima can temporarily disappear and the confined object is
transferred to the following minimum. A rocked ratchet may also operate without
thermal noise.

Phase transitions are usually characterized by an ``order parameter" that vanishes
for one of the phases and has a non zero value for the other phase. Near a second
order phase transition, the equilibrium size of the order parameter is small and can be comparable to that of its thermal fluctuations. The system we will consider is a 
superconducting loop near its transition temperature. Having a loop automatically 
provides the periodicity which
is encountered in Brownian motors and the presence of permanent currents in a
preferred direction might be thought of as a starting point for the required
asymmetric potential.

The feature that in my view turns a superconductor into an interesting
rectifying system is the presence of two competing currents that are governed
by independent fields: the supercurrent is governed by the ``order parameter",
whereas the normal current is governed by
the electromagnetic field. Both undergo thermal fluctuations. However, the 
strength of the electromagnetic fluctuations increases with temperature, 
whereas fluctuations of the order parameter are most important near the critical
temperature. This difference gives us the freedom to assign independent
sizes to each of these fluctuations.
In the system we will study, electromagnetic fluctuations will play the
role of ``Brownian motion", whereas fluctuations of the order parameter
will cause the time variation of the confining potential.

The simplest superconducting component we can consider is a Josephson 
junction. A Josephson junction is actually a zero-dimensional superconducting
wire: its state is completely described by the values of the order parameter
and of the electrochemical potential at its extremes. 
Therefore, in order to prevent the physical ideas from being obscured by
mathematical complexity, we will model our loop by a uniform superconducting 
wire interrupted by Josephson junctions. Modeling weak links by Josephson junctions will enable us to reduce the dynamics of the system to
standard textbook procedures.\cite{tink,likh,BP}
Ideas for using Josephson junctions as rectifying ratchets
have been studied in Refs.~[\onlinecite{Zapata,Falo,Weiss}]. Rectification of the 
motion of vortices has been considered in Ref.~\onlinecite{Nori1}.

In the following section we define the system to be considered and the
rules that govern its evolution. We also provide some motivation for the choices
involved in the model and some heuristics as to why rectification is expected to
occur. Section~\ref{RES} is a report of the numeric results. In Sec.~\ref{EXP} our 
model will be compared with available experiments. Finally, there is a summary
of the achieved results.

\section{OUR MODEL}
\subsection{Basic Features}
\begin{figure}
\scalebox{0.4}{\includegraphics{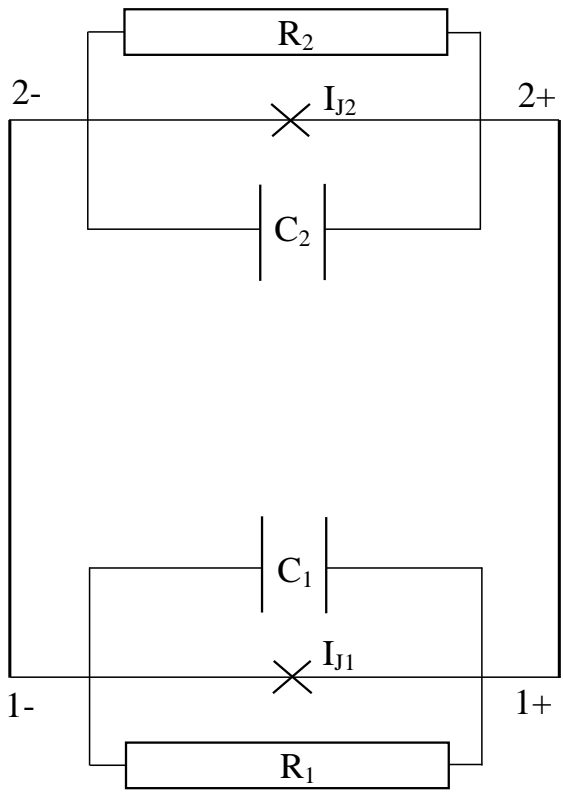}}%
\caption{\label{circuit}Schematic representation of the superconducting loop.
The vertical thick lines represent segments which are ``strongly" superconducting
and the horizontal branches are weak links, which are modeled by Josephson
junctions. These junctions are modeled by ``pure" junctions in parallel with
a resistor and a capacitor.}
\end{figure}

Figure \ref{circuit} shows schematically the system we will study.
$1\pm$ and $2\pm$ are points in the loop. Between $1+$ and $2+$ (resp. $-$)
there is a strong superconducting segment and between $1+$ and $1-$ (resp. 
$2$) there is a weak link. A weak link is a portion of a superconducting
wire where superconductivity is weaker, typically, a constriction. These 
weak links will be modeled by Josephson junctions.

The gauge-invariant phase difference across weak link $i$ is defined by
$\gamma_i=\varphi_{i+}-\varphi_{i-}+(2\pi/\Phi_0)\int_{i-}^{i+}{\bf A}
\cdot d{\bf s}$ and the electromotive force by
$V_i=\mu_{i+}-\mu_{i-}+(1/c)\int_{i-}^{i+}(\partial{\bf A}/\partial t)
\cdot d{\bf s}$, where $\varphi$ is the phase of the order parameter,
$\Phi_0$ the quantum of flux (contrary to the electron charge, $\Phi_0>0$), 
${\bf A}$ is the vector potential, $\mu$
the electrochemical potential, $c$ the
speed of light, $t$ the time, and the integrals are performed along the
link. For junction 1 the equilibrium value of the critical current
will be denoted by $\Ic$, the resistance by $R_1$, and the capacitance
by $C_1=C$. For junction 2 the equilibrium value of the critical current
will be denoted by $\alpha \Ic$ and,
in order to reduce the number of parameters in the model, the resistance
will be taken as $R_2=R_1/\alpha$ and the capacitance as $C_2=\alpha C$.
$\Ic$, $R_1$ and $\Phi_0/2\pi c=\hbar/2e$ will be taken as units; 
accordingly, the units of voltage, time, capacity and inductance are
$\Ic R_1$, $\hbar/2e\Ic R_1$, $\hbar/2e\Ic R_1^2$ and $\hbar/2e\Ic$,
respectively.
In these units the ac Josephson relation is
\be
d\gamma_i/dt=V_i \; .
\label{ac}
\ee

Additional equations are obtained by equating the current around the loop,
which is related to the magnetic flux encircled by the loop, with the
current flowing across each junction:
\be
I_{{\rm J}i}+V_i/R_i+C_i dV_i/dt+
I_{{\rm N}i}=-(-1)^i (\varphi_x+\gamma_2-\gamma_1)/L \; .
\label{current}
\ee
$I_{{\rm J}i}$ is the current that flows across the ``pure" junction, 
$V_i/R_i$ the current that ideally flows through the resistor according
to Ohm's law, $C_i dV_i/dt$ is the current that goes into the capacitor,
and $I_{{\rm N}i}$ is due to thermal fluctuations; $\varphi_x$
is the magnetic flux induced by external sources multiplied by $2\pi/\Phi_0$
and $L$ is the self inductance of the loop.

$I_{{\rm N}i}$ is the usual Johnson--Nyquist current.  
Let us choose a period of time
$\Delta t$ which is large compared to the autocorrelation time, but can be
treated as infinitesimal in macroscopic processes. Averaged over $\Delta t$,
$I_{{\rm N}i}$  has the form
\be
\bar I_{{\rm N}1}=\eta g_1/\sqrt{\Delta t};\;\;\;
\bar I_{{\rm N}2}=\eta g_2\sqrt{\alpha/\Delta t}\; ,
\label{noise}
\ee
with $\eta^2=2k_B T/R_1$, where $k_B$ is Boltzmann's constant, $T$ is the 
temperature, and $g_{1,2}$ are random
numbers with normal distribution, zero average and unit variance.
(This follows since 
$\langle(\bar I_{{\rm N}i}\Delta t)^2\rangle=\int_0^{\Delta t}dt'
\int_0^{\Delta t}dt''\langle I_{{\rm N}i}(t')I_{{\rm N}i}(t'')\rangle
\approx 2\Delta t\int_0^\infty\langle I_{{\rm N}i}(0)I_{{\rm N}i}(s)\rangle ds
=2k_B T\Delta t/R_i$.)

\subsection{Heuristic Considerations and Additional Choices}
If we ignore $I_{{\rm N}i}$, Eqs. (\ref{ac}) and (\ref{current}) predict an evolution 
for $\gamma_i$ that is equivalent to viscous motion down a potential 
$(\varphi_x+\gamma_2-\gamma_1)^2/2L+\mathcal{I}_1+\mathcal{I}_2$, where
$d\mathcal{I}_i/d\gamma_i=I_{Ji}$. In order to rectify Brownian motion, 
there should be some direction in the $\gamma_{1,2}$-space such that
this potential is periodic and is not symmetric under reflection. 

This symmetry breaking cannot be achieved if $I_{{\rm J}1}$ and $I_{{\rm J}2}$ 
are both sinusoidal functions of the phase
difference, as is the case for tunnel junctions. Fortunately, for superconducting
wires the current-phase dependence becomes sinusoidal only as a 
limit.\cite{Bara,Kulik,review} 
In general, $I_{{\rm J}i}$ has its maximum away from
$\gamma_i =\pi /2$. The deviation from sinusoidality can be quite large 
and $I_{{\rm J}i}$ can even be 
multivalued. Various types of nonsinusoidal current-phase relation were also
encountered in junctions such as superconductor-ferromagnet-superconductor
point contacts.\cite{NS}

\begin{figure}
\scalebox{0.85}{\includegraphics{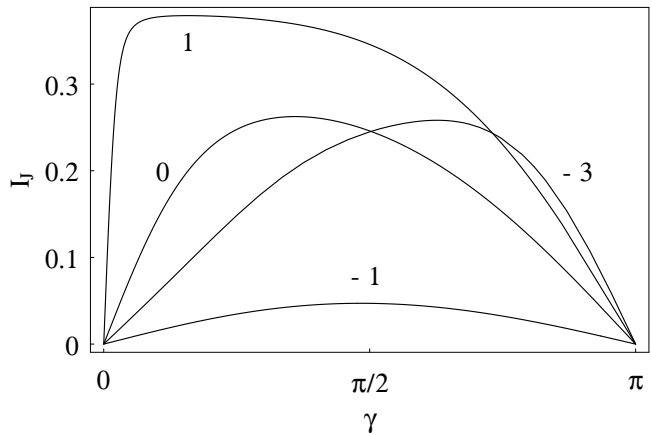}}%
\caption{\label{barat}Current through a Josephson junction as a function of the phase
difference for the case considered in Ref.~\onlinecite{Bara}. Here the unit of current
is $wc\xi_A H_c^2/\Phi_0$, where $w$ is the cross section of the wire, $\xi_A$ the
coherence length in the strongly superconducting wire and $H_c$ the thermodynamic
transition field. The length of the weak link is $2d\xi_A=0.002\xi_A$ and the curves
are for several values of $\Gamma$, a parameter that defines the ratio between the 
strengths of superconductivity in both materials. The curves are marked with the
number $\log_{10}(\Gamma/d)$. Only the curve for $\Gamma=10^{-1}d$ is nearly sinusoidal. 
In order to make the curve visible, the current values
for the case $\Gamma=10^{-3}d$ were multiplied by 400.}
\end{figure}

We may gain a better understanding of the present model by reviewing
Ref.~\onlinecite{Bara} in some detail. This reference considers a long and thin
wire $A$ with a region of it, denoted by $B$, replaced by a different material. 
$B$ is the ``weak link". Both materials are almost identical, the only difference
between them being that $B$ has a shorter mean free path. The weak link is 
characterized by two parameters. One of them, denoted by $\Gamma$, is the ratio
between Gorkov's universal function of the impurity parameter for both materials.
(Denoting the coherence lengths by $\xi_{A,B}$, $\Gamma=(\xi_B/\xi_A)^2$.) The
second parameter is the length of $B$, which is written as $2d\xi_A$. The article
provides explicit expressions that exactly solve the Ginzburg--Landau equations and
permit to evaluate the current-phase relation for the wire. We are interested
in temperature close to critical, so that we should consider $d\ll 1$; on the
other hand, any value $0<\Gamma<1$ is physically admissible. The authors conclude
that only in the regime $d^2\ll\Gamma\ll d$ the current-phase relation becomes 
sinusoidal. In order to appreciate the deviation from sinusoidality, Fig.~\ref{barat}
shows the current as a function of the phase difference for $d=0.001$ and several
values of $\Gamma$.

For simplicity, we shall keep only two harmonics 
and take
\be
I_{{\rm J}i}(\gamma_i)=I_{{\rm c}i}(\sin\gamma_i+\beta_i \sin 2\gamma_i)\; ,
\label{beta}
\ee
where $I_{{\rm c}i}$ and $\beta_i$ are constants that characterize the
junction $i$.
In order to present results that are clearly visible, most of our figures
use the values $\beta_1=-\beta_2=0.7$, but let me emphasize that the numerical
investigation covered the range $0\le|\beta_i|\le 1$ and the qualitative
conclusions of the following section remain valid for small values of
$|\beta_i|$.

In Ref.~\onlinecite{Zapata} the second harmonic is brought in by connecting two 
junctions in series. It should be noted, though, that this method involves
the implicit assumption that the same current flows through both ``pure" junctions
in series; this assumption is not acceptable in the context of this article, 
since the currents through both resistors fluctuate independently.

\begin{figure}
\scalebox{0.85}{\includegraphics{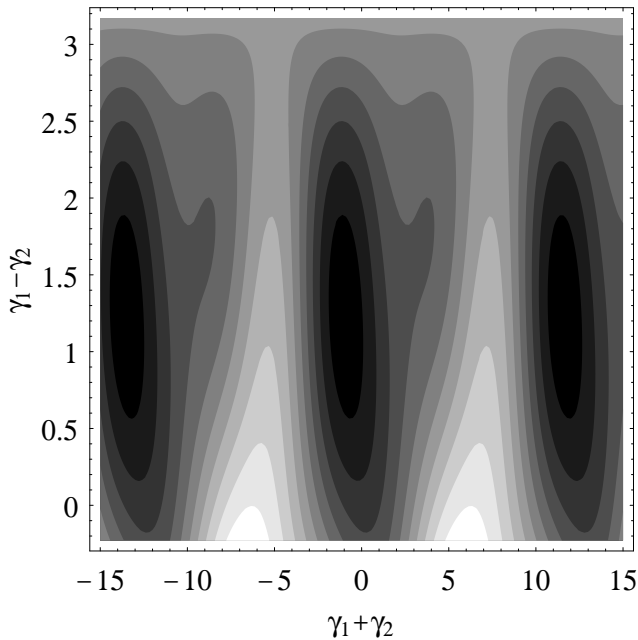}}%
\caption{\label{contour}The values of $\gamma_i$ behave as the coordinates of
a particle in two dimensions (with anisotropic mass) which feels the potential in 
this graph, in addition
to an anisotropic viscosity and a Langevin force. In this graph the darker
areas represent lower potentials. This potential is periodic in $\gamma_1+\gamma_2$,
with period $4\pi$. The parameters used to evaluate this graph
are $L=I_{{\rm c}1}=I_{{\rm c}2}=1$, $\beta_1=-\beta_2=0.7$, $\varphi_x=\pi/2$.}
\end{figure}

Using Eq.~(\ref{beta}), the ``potential" that describes the evolution of
$\gamma_i$ becomes 
$(\varphi_x+\gamma_2-\gamma_1)^2/2L 
-I_{{\rm c}1}(\cos\gamma_1+\frac{1}{2}\beta_1 \cos2\gamma_1)
-I_{{\rm c}2}(\cos\gamma_2+\frac{1}{2}\beta_2 \cos2\gamma_2)$.
A contour graph of this potential is shown in
Fig.~\ref{contour} for a given set of
parameters. The dark areas are minima, and the light areas are barriers
between them. If the barriers are high, the values of $\gamma_i$ will be 
confined within a minimum, but, if the barriers are lowered, the noise will cause
these values to diffuse, either to the right or to the left. Since the barrier
at the left is closer than that at the right, the probability of crossing it
is larger than that of crossing the barrier at the right. Therefore, if the barriers
are raised again after a suitable time, the values of $\gamma_i$ may be pushed
towards the consecutive minimum at the left, with a greater probability 
than that of being pushed to the right. In this way we may expect a non zero
average drift of $\gamma_1+\gamma_2$; $\gamma_1-\gamma_2$ remains always close to
$\varphi_x$.

We require a mechanism to vary the asymmetric potential of Fig.~\ref{contour} in 
time. This is provided by fluctuations of $I_{{\rm c}1}$ and $I_{{\rm c}2}$.
The maximum current through a junction is proportional to $|\psi|^2$, where 
$\psi$ is the order parameter. The evolution of $\psi$ can be described by the
time-dependent Ginzburg--Landau theory with the addition of a Langevin 
``force".\cite{Schmid} The ``standard" terms drive $\psi$ towards equilibrium,
whereas the Langevin term causes stochastic deviations from it. Since the 
Langevin term increases with the absolute temperature and the ``standard" terms 
increase with the distance from the critical temperature, the influence of
fluctuations is largest close to the critical temperature.

Due to the evolution of $\psi$, the evolution of $I_{{\rm c}i}$ should also
have the form of a random walk,
together with relaxation to the equilibrium value with some characteristic time
$\tau_i$. The explicit implementation we took was
\be
I_{{\rm c}1}(t+\Delta t)=\frac{\tau_1 \left[I_{{\rm c}1}(t)+(\delta_0 r_0+
\delta_1 r_1)\sqrt{\Delta t}\right]+ \Delta t \Ic}{\tau_1+ \Delta t}
\label{delta}
\ee
and, similarly, for $I_{{\rm c}2}$ the index 1 is replaced by 2, $\Ic$ by 
$\alpha\Ic$ and $\delta_0$ by $\alpha\delta_0$. Here $\Delta t$ is the short 
period of time we have chosen, $r_0$, $r_1$ and $r_2$ are random numbers distributed 
between $-1$ and $1$ and the $\delta_i$'s denote the strength of the fluctuations.
If $I_{{\rm c}1}$ or $I_{{\rm c}2}$ becomes negative, it is reset to 0.
$\delta_0$ describes synchronous fluctuations of both junctions, whereas
$\delta_{1,2}$ describe independent fluctuations.
If the coherence length is larger than the perimeter of the loop, as is the case
for mesoscopic loops near the critical temperature, we may expect that
$\delta_0$ will dominate, and vice versa. 

If the $\delta_i$'s are very small, Eq.~(\ref{delta}) reduces to exponential
relaxation of $I_{{\rm c}i}$ to its equilibrium value, with time constant $\tau_i$;
in the opposite extreme, if the relaxation time becomes infinite, we are left 
with pure random walk.

In order to reduce the number of 
parameters in the model we took $\beta_1=-\beta_2=\beta$, $\tau_1=\tau_2=\tau$
and $\delta_2=\alpha\delta_1$. We are still left with quite a large number of
parameters, but we shall see that several parameters have just a minor influence
on the results; in addition, several parameters can be grouped, in the sense that
varying any of them has qualitatively the same effect.

The relaxation times and the size of the random walk steps are related by the
fluctuation--dissipation theorem. For example, let us consider the case that
$\delta_0=0$ and the part of the energy of junction 1 that depends on $I_{{\rm c}1}$
has the form $\kappa(I_{{\rm c}1}-\Ic)^2$. The constants $\kappa$ and $\Ic$
could be obtained, for instance, from the Ginzburg--Landau theory; both depend on the
fabrication details of the junction. In addition, $\Ic$ should be proportional to the
distance to the critical temperature and $\kappa$ should be independent of the temperature. Let $F$ be the ``Langevin force" that acts on $I_{{\rm c}1}$. Then, by standard methods,
we obtain $\int_{-\infty}^{\infty}\langle F(0)F(s)\rangle ds=k_BT/\kappa\tau_1$.
From here, it can be obtained that the average of the square of the change
of $I_{{\rm c}1}$ during the period $\Delta t$ will be $k_BT\Delta t/\kappa\tau_1$.
It then follows that $\tau_1\delta_1^2=2k_BT/\kappa$. However, since $\kappa$
depends on the details of the juction, we will regard $\tau$ and the
$\delta_i$'s as independent parameters.

The precise form of the fluctuations of $I_{ci}$ is not important. I have
changed the distributions of the random numbers $r_i$ from uniform to Gaussian,
to bimodal and to asymmetric (retaining the values of the average and the
variance) and the difference in the results obtained from different distributions
is not noticeably larger than the typical difference obtained for different runs
with the same distribution. All that matters is how often and for how long
the potential barriers in Fig.~\ref{contour} are low compared to the noise level.
The scattering in the results, though, seems to depend on the distribution used.

One might gain some intuition by comparing the orders of magnitude of the energies
involved in the problem. Fluctuations in the electromagnetic field and in the
order parameter are both of the order of $k_BT$; the energy required to break
superconductivity ($\sim\kappa\Ic^2$) should be significantly, but not exceedingly,
larger than $k_BT$. In this way, only when many random steps accumulate to
diminish the order parameter, superconductivity is broken at a junction. For the
parameters used here, $I_{{\rm c}i}$ typically vanished in one out of $10^4$
steps.

\section{RESULTS}
\label{RES}
The equations above determine the evolution of the phase differences 
$\gamma_{1,2}$ for given sets of model parameters. This evolution was followed, 
using Euler iterations, during
$10^9$ steps. The average voltages are then obtained by dividing the change
of $\gamma_{1,2}$ by the elapsed time. In order to cancel out possible biases in
the random numbers, we evaluated four sets of phase differences, which were
obtained by reversing the sign of either $\bar I_{{\rm N}1}$ or $\bar I_{{\rm N}2}$ 
in Eq.~(\ref{noise}), and then took the average of the voltages obtained for
the four sets. The values obtained for $V_1$ and $V_2$ were nearly the same,
and the reported voltage values are their averages.

\begin{figure}
\scalebox{0.85}{\includegraphics{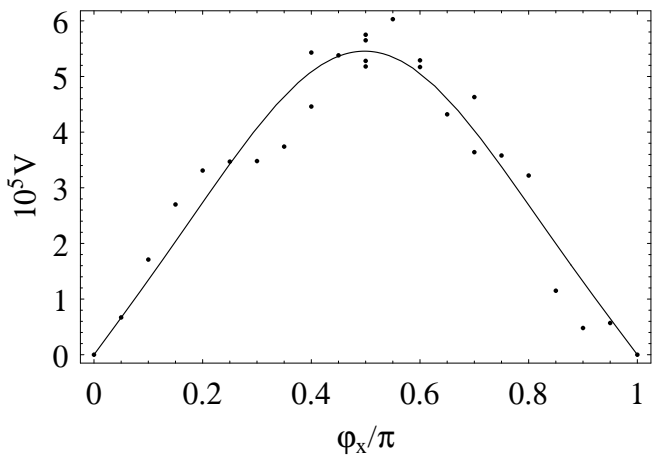}}%
\caption{\label{fix}Voltage as a function of the magnetic flux
$\varphi_x/\pi=2\Phi/\Phi_0$. (This function is odd, and periodic in $\Phi$
with period $\Phi_0$.)
The parameters used here were $\alpha=1$, $C=0$, $L=1$, $\beta=-0.7$, $\delta_0=0.02$, 
$\delta_1=0$, $\tau=5\times 10^4$, $\eta=0.45$. For the dependence of the voltage
on the parameters, see text. The curve is a guide for the eye.}
\end{figure}

Figure \ref{fix} shows our main result, obtained for a given set of parameters.
In spite of the scattering in the results, it can be safely concluded that the
dc voltage does not vanish, that it is a function of the flux, that it vanishes
for integer and half-integer values of $\Phi/\Phi_0$, and has maxima in between.

\begin{figure}
\scalebox{0.85}{\includegraphics{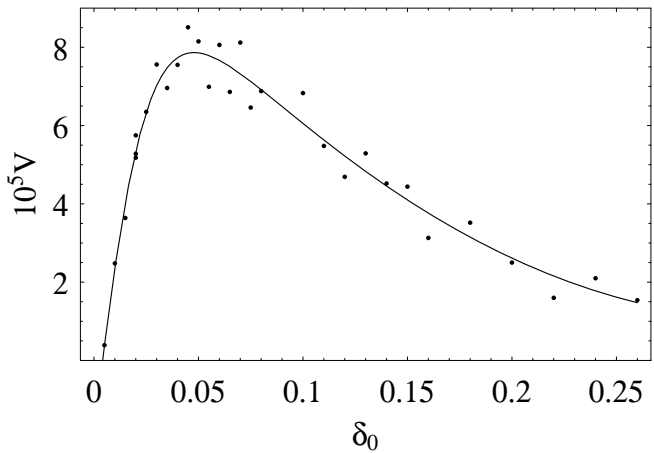}}%
\caption{\label{fluct}Voltage as a function of the strength of the fluctuations
of the maximal supercurrent of the junctions. The flux is fixed at $\varphi_x=0.5\pi$
and all the other parameters (except for $\delta_0$) are the same as in Fig.~\ref{fix}.
The curve is a guide for the eye.}
\end{figure}

I have studied the dependence of the voltage on each of the parameters of the 
model. This study has been limited to variations of a single parameter each
time, with the others fixed at their values as in Fig.~\ref{fix}. For $\delta_0$,
$\tau$ and $\eta$, the voltage is insignificant below some threshold, then rises
to a maximum and decreases slowly. This is
expected, because for low values of these parameters the phase differences have
no opportunities to jump between consecutive minima of the ``potential", whereas
for high values there are too many opportunities and thermal equilibrium is reached.
Figure~\ref{fluct} shows the voltage as a function of $\delta_0$. 
Qualitatively, the same shape is obtained for quantities playing equivalent roles
in diverse models, e.g. in Fig.~2 of Ref.~\onlinecite{charles}. As a function of $\eta$,
the voltage attains half of its maximum value at $\eta\sim 0.3$ and $\eta\sim 0.8$.
As a function of $\tau$, the voltage rises sharply to a maximum near $\tau=2\times 10^4$
and then decreases very slowly. The reason is that the typical distances 
of $I_{{\rm c}1}$ and $I_{{\rm c}2}$ from their equilibrium values are actually not proportional to $\tau$, but rather to its square root.

The dependence of the voltage on the other parameters is roughly as follows.
It has a maximum close to $\alpha\sim 1.1$ and decreases to half value for
$\alpha\sim 1/3$ and $\alpha\sim 4$. It is insensitive to $C$ for $C\alt 1$
and decreases to half value for $C\sim 10^3$. It is also insensitive to $L$ for $L\alt 1$
and decreases to half value for $L\sim 12$. For $0\le|\beta|\le 1$ the voltage is roughly
proportional to $-\beta+0.25\beta^3$ and, for $\delta_1\le 0.03$, to $e^{-80\delta_1}$.

In principle, the results should not depend on $\Delta t$, but in practice the
steps cannot be too large if the derivatives of $\gamma_i$ and $V_i$ are approximated by 
ratios of discrete differences; $\Delta t$ also cannot be too small, since then too many
steps are required until $I_{{\rm c}1,2}$ become small a statistically significant number
of times. In our calculations we took $\Delta t=0.01$. In order to estimate the
accuracy of our results, we repeated some calculations for $\Delta t=0.1$. The
voltages obtained were lower by about 20\%.

An important difference between our model and other flashing ratchets is that 
here the variations of $I_{{\rm c}i}$ in time are themselves due to spontaneous
fluctuations, and it is not obvious who is the agent that invests the work required
for rectification. Some models have been studied in which rectification is 
performed by fluctuations,\cite{charles,bio} but those are manifestly nonequilibrium
fluctuations, whereas our heuristic considerations suggest that in the present case
rectification is not subject to any detailed form of $\bar I_N$ or of the
fluctuations of $I_{{\rm c}i}$, so that the question remains open.

\section{APPLICATION TO EXISTENT EXPERIMENT}
\label{EXP}
Recent measurements on mesoscopic rings composed by segments of unequal widths, near
the critical temperature,\cite{nikul} found a dc voltage with apparently the same
flux-dependence as in our Fig.~\ref{fix}. Thus far we have not explained their results,
since the voltages we have found are smaller than $\Ic R_1$ by four orders of magnitude,
whereas the experimental results are smaller than the corresponding product by only
one order of magnitude (provided that we identify $R_i$ with the resistance of a
segment in its normal state and $I_{{\rm c}i}$ with the maximum supercurrent that
can flow in it). 

Nevertheless, bearing in mind the differences between the
experiment and our model, the very existence of a non zero flux-dependent
dc voltage in our case is already remarkable. The experiment differs from our model
in two essential respects. The first difference is that their rings contain only two 
segments, without the strongly superconducting branches (see Fig.~\ref{circuit}).
Conceivably, by assuming that $I_{{\rm J}i}$ depends only on $\gamma_i$ and by 
assuming that fluctuations affect the size of $I_{{\rm J}i}$ only (with the shape of 
the functions remaining invariant), essential features
of the ring behavior are lost.

The second difference between the experiment and our model is that the experiment
used an external ac current $I_x(t)$ to destroy superconductivity. This feature is
readily incorporated into our model. Assuming that both segments give equal 
contributions to the self inductance of the ring, all we have to do is add $I_x(t)/2$ 
at the right hand side of Eq.~(\ref{current}). By adding this term, our system becomes
a rocked ratchet.\cite{omega,Mag}

\begin{figure}
\scalebox{0.85}{\includegraphics{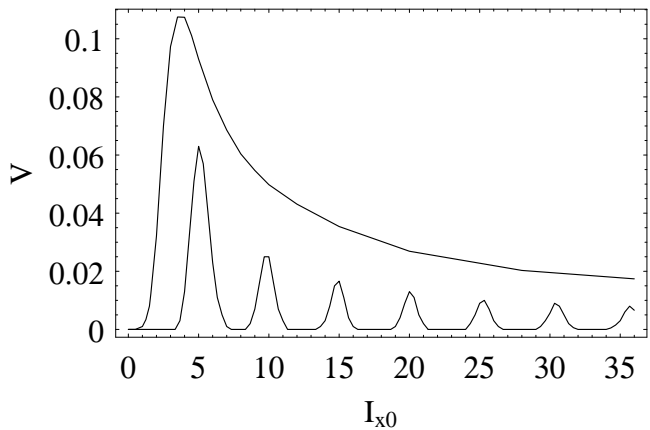}}%
\caption{\label{Ix}Average voltage as a function of the amplitude of the external
ac current. The upper curve was evaluated for a period $2\pi/\omega=5\times 10^4$,
but we also searched for periods in the range between 50 and $5\times 10^6$, and
no appreciable difference was found; the lower curve is for $2\pi/\omega=10$.
The other parameters used here were $\alpha=2$, $C=0$, $L=0.03$, $\beta=0.3$, $\delta_0=0.02$, 
$\delta_1=0$, $\tau=800$, $\eta=0.45$ and $\varphi_x=\pi/2$.}
\end{figure}

Figure~\ref{Ix} shows the dc voltages obtained for our model with an external current 
of the form $I_x(t)=I_{x0}\sin(\omega t)$. 
Some of the parameters used to evaluate these results ($\alpha$, $C$, $L$) were
estimated from the available experimental data; the remaining parameters are 
arbitrary. The results bear close resemblance to
those obtained for other periodically rocked ratchets:\cite{Weiss,omega} for rocking periods
$2\pi/\omega$ that are larger by two or more orders of magnitude than the characteristic time $\hbar/2e\Ic R_1$ of the circuit, we obtain the upper curve, independently
of $\omega$. However, for periods that are comparable to $\hbar/2e\Ic R_1$, we find
that there is a nearly periodic pattern of rocking strengths for which the rectifying
effect is present.

In the experiment, $\hbar/2e\Ic R_1\sim 10^{-11}$s, whereas the shortest period
of the applied ac current was $10^{-6}$s. The relevant curve is therefore the
upper one. Indeed, the experiment found that the frequency of the ac current has no 
influence. The upper curve in Fig.~\ref{Ix} is remarkably similar to Fig.~6 in 
Ref.~\onlinecite{nikul}, and this time the orders of magnitude coincide (for the voltages,
and also for the currents). 

The heuristic argument used in Fig.~\ref{contour} to predict the existence of a
dc voltage is not applicable now: in the presence of a large external current,
the sign of the voltage is opposite to what that argument would predict. 

Even in the presence of an external current, the noise and the
supercurrent fluctuations play a central role. For $I_{x0}=4$ and the other parameters
as in the upper curve of Fig.~\ref{Ix}, setting $\eta=\delta_0=0$ leads to a voltage decrease by one order of magnitude; for $I_{x0}\alt 2.1$, the voltage vanishes when we set
$\eta=\delta_0=0$.

The experimental temperature is related to the parameters in our model through the
values of $\Ic$. The typical deviation of $I_{{\rm c}i}$ from its equilibrium value 
is of the order of $\delta_i\tau^{1/2}\sim(k_BT/\kappa)^{1/2}$. A significant
response is obtained when $\Ic$ is larger than---but comparable to---this typical
deviation, i.e., for temperatures slightly below the onset of superconductivity,
as in this experiment.

Let us finally discuss two additional experiments in which nonuniform superconducting loops
lead to a flux dependent dc voltage. Long ago de Waele et al.\cite{Bruyn} measured
dc voltage on a double point contact SQUID. The asymmetry was not in the weak links,
as in the present article, but rather in the thick parts of the loop: one part was
a niobium foil and the other a tin needle. No controlled driving ac current was supplied 
to the circuit; apparently, the role of the ac current was substituted by existing
electromagnetic radiation in the lab. When the circuit was appropriately shielded,
the dc voltage was no longer measurable. As in the cases considered here, the highest
dc voltage was found when the temperature was slightly lower than the critical
temperature of the weaker superconducting material (Sn). The highest dc voltage 
was comparable to the product of the maximal current through a point contact times its normal resistance.

Weiss et al.\cite{Weiss} measured rectification by an YBCO SQUID for a wide range
of applied frequencies (ac and rf). The asymmetry was mainly due to different
maximal currents of the junctions; this asymmetry leads to a rocking ratchet.
Their ac results, shown in their Fig.~5(a) are remarkably similar to the upper curve
in our Fig.~\ref{Ix}, including orders of magnitude. In later experiments\cite{applied}
the same group studied the influence of temperature and repeated the experiment
for a niobium circuit. However, all the temperatures were considerably below the
critical temperature, so that only variations in our parameter $\eta$ were significant, while the fluctuations of $I_{{\rm c}i}$ had no noticeable influence. In addition, they used tunnel
junctions, so that our $\beta$ is also expected to vanish.

In view of the differences among the reviewed experimental loops and our ``zero-dimensional"
model, their similar behavior is remarkable and suggests that the features encountered
in our results will be present in any sort of nonuniform superconducting loop near its
transition temperature.

\section{SUMMARY}
\label{FIN}
Superconductors near their transition temperature may be regarded as a special class
of thermal ratchets, in the sense that the fluctuations of its order parameter can act
as a flashing asymmetric ``potential" and thus rectify the Johnson noise.
We have found a simple model that qualitatively reproduces the
experimental results of Ref.~\onlinecite{nikul}. For applied frequencies that are
comparable to the ``natural" frequency of the circuit, a complex pattern is
predicted.
In the absence of external current, the voltage decreases by about three orders
of magnitude and may change sign, but our simulations and our heuristic considerations
indicate that it does not vanish.

\begin{acknowledgments}
This work has been supported in part by the Israel Science Foundation. I wish
to thank Konstantin Likharev and Howard Wiseman for their answers to my inquiries.
I am also indebted to Peter H\"{a}nggi and Jacob Rubinstein for reading early
versions of the manuscript.
\end{acknowledgments}


\begin{thebibliography}{9}
\bibitem{astum1} R. D. Astumian and P. H\"{a}nggi, Physics Today, November
2002, p. 33.
\bibitem{Rei}P. Reimann, Phys. Rep. {\bf 361}, 57 (2002).
\bibitem{astum2} R. D. Astumian, Science {\bf 276}, 917 (1997).
\bibitem{RMP}F. J\"{u}licher, A. Adjari and J. Prost, Rev. Mod. Phys. {\bf 69}, 1269 
(1997).
\bibitem{Zim}P. H\"{a}nggi and R. Bartussek in {\it Nonlinear Physics of Complex
Systems}, Lect. Notes Phys. Vol. {\bf 476} edited by J. Parisi, S. C. M\"{u}ller
and W. Zimmermann (Springer, Berlin, 1996) pp. 294-308. 
\bibitem{exp}L. P. Faucheux, L. S. Bourdieu, P. D. Kaplan, and
A. J. Libchaber, Phys. Rev. Lett. {\bf 74}, 1504 (1995);
J. Rousselet, L. Salome, A. Adjari, and J. Prost,
Nature {\bf 370}, 446 (1994); K. Svoboda, C. F. Schmidt, B. J. Schnapp, and
S. M. Block, Nature {\bf 365}, 721 (1993).
\bibitem{tink} M. Tinkham, {\it Introduction to Superconductivity} 
(McGraw-Hill, New York, 1996).
\bibitem{likh} K. K. Likharev, {\it Dynamics of Josephson Junctions and
Circuits} (Gordon and Breach, New York, 1986).
\bibitem{BP}A. Barone and G. Patern\`{o}, {\it Physics and Applications of
the Josephson Effect} (Wiley, New York, 1982).
\bibitem{Zapata}I. Zapata, R. Bartussek, F. Sols, and P. H\"{a}nggi,
Phys. Rev. Lett. {\bf 77}, 2292 (1996).
\bibitem{Falo}F. Falo, P. J. Martinez, J. J. Mazo, and S. Cilla,
Europhys. Lett. {\bf 45}, 700 (1999).
\bibitem{Weiss} S. Weiss, D. Koelle, J. M\"uller, R. Gross, and K. Barthel,
Europhys. Lett. {\bf 51}, 499 (2000). 
\bibitem{Nori1}J. F. Wambaugh, C. Reichhardt, C. J. Olson, F. Marchesoni, and
F. Nori, Phys. Rev. Lett. {\bf 83}, 5106 (1999);
C. J. Olson, C. Reichhardt, B. Janko, and 
F. Nori, Phys. Rev. Lett. {\bf 87}, 177002 (2001);
J. E. Villegas, S. Savel'ev, F. Nori, E. M. Gonzalez, J. V. Anguita, 
R. Garc\'{i}a, and J. L. Vicent, Science {\bf 302}, 1188 (2003).
\bibitem{Bara}A. Baratoff, J. A. Blackburn and B. B. Schwartz, Phys. Rev. Lett.
{\bf 25}, 1096 (1970); Phys. Rev. Lett. {\bf 25}, 1738 (1970)(errata).
\bibitem{Kulik}A. A. Golubov, M. Yu. Kupriyanov, and E. Il'ichev, Rev. Mod. Phys. 
{\bf 76}, 411 (2004).
\bibitem{review} K. K. Likharev, Rev. Mod. Phys. {\bf 51}, 101 (1979).
\bibitem{NS} A. A. Golubov, M. Yu. Kupriyanov, and Ya. V. Fominov,
Pis'ma Zh. Eksp. Teor. Fiz.
{\bf 75}, 709 (2002) [JETP Lett. {\bf 75}, 588 (2002)].
\bibitem{Schmid} A. Schmid, Phys. Rev. {\bf 180}, 527 (1969).
\bibitem{charles}C. R. Doering, W. Horsthemke, and J. Riordan,
Phys. Rev. Lett. {\bf 72}, 2984 (1994).
\bibitem{bio}R. D. Astumian and M. Bier, Biophys. J. {\bf 70}, 637 (1996).
\bibitem{nikul}S. V. Dubonos, V. I. Kuznetsov, I. N. Zhilyaev, A. V. Nikulov,
and A. A. Firsov, Zh. Eksp. Teor. Fiz. Pis'ma Red. {\bf 77}, 439 (2003) 
[JETP Lett. {\bf 77}, 371 (2003)]. 
\bibitem{omega}R. Bartussek, P. H\"{a}nggi and J. G. Kissner, Europhys.
Lett. {\bf 28}, 459 (1994).
\bibitem{Mag}M. O. Magnasco, Phys. Rev. Lett. {\bf 71}, 1477 (1993).
\bibitem{Bruyn} A. Th. A. M. de Waele, W. H. Kraan, R. de Bruyn Ouboter,
and K. W. Taconis, Physica {\bf 37}, 114 (1967).
\bibitem{applied}A. Sterk, S. Weiss, and D. Koelle,
Appl. Phys. A {\bf 75}, 253 (2002).
\end{thebibliography}
\end{document}